\documentstyle[epsf]{article}

\newcommand{\bec}{Bose--Einstein condensate}

\def\be{\begin{equation}}
\def\ee{\end{equation}}
\def\bea{\begin{eqnarray}}
\def\eea{\end{eqnarray}}

\begin{document}
\baselineskip=4.2mm  
\twocolumn[\hsize\textwidth\columnwidth\hsize\csname @twocolumnfalse\endcsname

\begin{centering}

{\Large \bf Spin domains in ground state spinor Bose-Einstein
condensates  \\[2ex]}

{J. Stenger, S. Inouye, D.M. Stamper-Kurn, H.-J. Miesner, A.P.
Chikkatur, and W. Ketterle}

{\it Department of Physics and Research Laboratory of
Electronics, \\
Massachusetts Institute of Technology, Cambridge, MA 02139}

\vspace*{1.3cm}


\end{centering}
]
\vspace{2cm}

{\bf Bose--Einstein condensates of dilute atomic gases,
characterized by a macroscopic population of the quantum mechanical
ground state, are a new, weakly
interacting quantum fluid \cite{ande:95,davi:95,brad:97}.
In most experiments condensates in a single weak field
seeking state are magnetically trapped. These condensates can be described
by a scalar order parameter similar to the spinless superfluid $^4$He. Even
though alkali atoms have angular momentum, the spin orientation is not a
degree of freedom because spin flips lead to untrapped states and are
therefore a loss process. In contrast, 
the recently realized 
optical trap for sodium condensates \cite{stam:98} confines
atoms independently of their
spin orientation. This opens the possibility to study spinor condensates
which represent a
system with a vector order parameter instead of a scalar.
Here we report a study of the equilibrium state of spinor
condensates in an optical trap. The freedom of spin orientation leads to the
formation of spin domains in an external magnetic field. The structure of
these domains are
illustrated in spin domain diagrams. Combinations of
both miscible and immiscible spin components were realized.  }

A variety of new phenomena is predicted \cite{ho:98,ohmi:98,law:98}
for spinor condensates, such
as spin textures, propagation of spin waves and coupling between superfluid
flow and atomic spin. To date such
effects could only be studied in superfluid $^3$He, which can be described
by Bose--Einstein condensation of Cooper pairs of quasi particles having
both spin and orbital angular momentum \cite{voll:90}. Compared to the
strongly interacting $^3$He, the properties of weakly interacting \bec s of
alkali gases can be calculated by mean field theories in a much more
straightforward and simple way.

Other systems which go beyond the description with a single scalar order
parameter are condensates of two different hyperfine states of $^{87}$Rb
confined in magnetic traps. Recent experimental studies have explored the
spatial separation of the two components \cite{myat:97,hall:98a} and their
relative phase \cite{hall:98b}.
Several theoretical papers describe their 
structure \cite{sigg:80,ho:96,timm:97,esry:97,ohbe:98,pubi:98,ao:98} and their collective
excitations \cite{busc:97,grah:98,pu:98,esry:98}.

Compared to these two--component condensates, spinor condensates have
several new features including the vector character of the
order parameter and the changed role of spin relaxation collisions which
allow for population exchange among hyperfine states without trap loss. In
contrast, for $^{87}$Rb experiments trap loss due to spin relaxation
severely limits the lifetime. 

We consider an $F\!=\!1$ spinor condensate subject to spin relaxation, in
which two $m_F\!=\!0$ atoms can collide and produce an $m_F\!=\!+1$ and an
$m_F\!=\!-1$ atom and vice versa. We investigate the distribution of
hyperfine states and the spatial distribution in equilibrium assuming
conservation of the total spin.

The ground state spinor wave function is found by minimizing the free
energy \cite{ho:98}
\be
K  =  \int \,d^3r\,n\left[
     V + \frac{c_0n}{2}   +\frac{c_2n}{2}
	\langle\vec{F}\rangle ^2  + E_{ze} - p_0 \langle F_z\rangle \right], \label{k}
\ee

\noindent where kinetic energy terms are neglected in the Thomas-Fermi
approximation which is valid as long as the dimension of spin domains 
(typically 50 $\mu m$) is larger
then the penetration depth \cite{ao:98} (typically 1 $\mu m$). 
$V$ is the trapping potential, $n$ is the density, $\vec{F}$
is the angular momentum per atom, and $E_{ze}$ is the Zeeman energy in an
external magnetic field. The Lagrange multiplier $p_0$ accounts for the
total spin conservation. The mean field energy in Eqn. (\ref{k}) consists
of a spin independent part proportional to $c_0$ and a spin dependent part
proportional to $c_2\langle\vec{F}\rangle^2$. The coefficients $c_0$ and
$c_2$ are related to the scattering lengths $a_0$ and $a_2$ for two
colliding atoms with total angular momentum $F_{tot}=0$ or $F_{tot}=2$ by
$c_0  =  4\pi\hbar^2\bar{a}/M $ and $c_2  =  4\pi\hbar^2\Delta a/M $ with
$\bar{a}=(2a_2 + a_0)/3 $, $ \Delta a=  (a_2 - a_0)/3$, and $M$ for the
atomic mass \cite{ho:98}. The spin dependent interaction originates from 
the term $c_2 \vec{F_1}\cdot\vec{F_2}$
in the interaction of two atoms, which is ferromagnetic for $c_2<0$ and 
anti--ferromagnetic for $c_2>0$.

In the Bogoliubov approach the many--body ground state wave function is
represented by the spinor wave function
\be
\Psi (\vec{r})= \sqrt{n(\vec{r})} \;\zeta (\vec{r}) = \sqrt{n(\vec{r})}\;
	\left( \zeta_+(\vec{r}), \zeta_0(\vec{r}), \zeta_-(\vec{r}) \right) ,
\label{spinor}
\ee

\noindent where $\zeta_+, \zeta_0, \zeta_-$ denote the amplitudes for the
$m_F\!= +1, 0, -1$ states, respectively, and $|\zeta |^2=1$.

The Zeeman energy $E_{ze}$ is given by
\be
E_{ze} = E_+ |\zeta_+|^2 + E_0 |\zeta_{0}|^2 + E_- |\zeta_-|^2
 = E_0 - \tilde{p}\, \langle F_z\rangle + q\, \langle F_z^2\rangle .
 \ee

\noindent $E_+, E_0, E_-$ are the Zeeman energies of the $m_F=+1,0,-1$
states, $2q\equiv E_+ + E_- - 2 E_0$ is the Zeeman energy difference in a
spinflip collision, and $2\tilde{p} \equiv E_- - E_+$.  The $E_0$ term can
be included in the trapping potential $V$. The parameter $\tilde{p}$ can be
combined with the Lagrange multiplier $p_0$  to give $p\equiv
\tilde{p}+p_0$.

In the following we determine the spinor which minimizes the
spin--dependent part $K_s$ of the free energy:
\be
K_s = c\, \langle \vec{F}\rangle ^2 - p\, \langle F_z\rangle + q\, \langle
F_z^2\rangle , \label{kint}
\ee

\noindent where $c=c_2 n/2$.
The minimization of Eqn. (\ref{kint}) for different values of the
parameters c, p, and q is straightforward, and is shown graphically in the
form of spin-domain diagrams in Fig. 1. 

Experimentally, the values of c, p, and q can be varied arbitrarily,
representing any region of the spin domain diagram.
The magnitude (but not the sign) of the coefficient
$c$ is varied by changing the density $n$, either by changing the
trapping potential, or by studying condensates with different numbers of atoms.
In this study, the axial length of the trapped condensate is more than 60
times larger than its radial size, and thus we consider the system
one--dimensional, and integrate over the radial coordinates, obtaining
$n = 2 n_0/3$ where $n_0$ is the density at the radial center.
This integration assumes a parabolic density profile within the
Thomas--Fermi approximation. 
The value of $q$ can be changed by applying a weak external bias field $B_0$;
$q$ then corresponds to the quadratic Zeeman shift $q = \hat{q} B_0^2$.
The coefficient $p$ arises both from the linear Zeeman shift and from the
Lagrange multiplier $p_0$ which is determined by the total
spin of the system. For a
system with zero total spin in a homogenous bias field $B_0$, $p_0$ cancels
the linear Zeeman shift due to $B_0$, yielding $p = 0$.
Positive (negative) values of $p$ are achieved for condensates with a
positive (negative) overall spin.
Finally, the coefficients can be made to vary spatially across the condensate.
In particular, applying a field gradient $B'$ along the axis of the trapped
condensate causes $p$ to vary along the condensate length.
For a condensate with zero total spin,
$p = \mu B' z$ where $z$ is the axial coordinate
with $z = 0$ at the center of the condensate.
Thus, the condensate samples a vertical line in the
spin domain diagrams of Fig. 1.  The center of this line lies at $p = 0$, and
its length is given by the condensate length scaled by $\mu B'$.

The experimental study of spinor condensates required techniques to
selectively prepare and probe condensates in arbitrary hyperfine states.
Spinor condensates were prepared in several steps. Laser cooling and
evaporative cooling were used to produce sodium condensates in the $m_F\!
=\! -1$ state in a cloverleaf magnetic trap \cite{mewe:96}. The condensates
were then transfered into an optical dipole trap consisting of a single
focused infrared laser beam \cite{stam:98}. 
Arbitrary populations of the three hyperfine states were prepared using rf
transitions \cite{stam:98}. 
After the spin preparation, a bias field $B_0$ and a field gradient
$B^\prime$ were applied for a variable amount of time (as long as 30 s),
during which the atoms relaxed towards their equilibrium distribution, 
as shown in fig. 2.

The profiles in Fig. 3 were obtained from vertical cuts through 
absorption images. They provide clear evidence of 
anti--ferromagnetic interaction. 
The spin structure is consistent with the corresponding spin domain diagram in 
fig. 1 {\bf a}.
Overlapping $m_F\! =\! \pm 1$ clouds as observed are incompatible
with the assumption of ferromagnetic interaction.

The strength $c=(50\pm 20)\,{\rm Hz}$ of the anti--ferromagnetic interaction 
was estimated by determining $z_b$, the location of the $m_F\!=\!0$ to the
$m_F\!=\! \pm 1$ boundary, and by plotting $p = \mu B^\prime z_b$ versus the
quadratic Zeeman shift $q=\hat{q}B_0^2$ as shown in Fig. 4. With
$n=(2.9\pm 0.5)\times 10^{14}\,{\rm cm}^{-3}$ the difference between
the scattering lengths can be determined to
$a_2-a_0 = 3\Delta a = (3.5\pm 1.5) a_B=(0.19\pm 0.08)\,{\rm nm}$
where $a_B$ denotes the Bohr radius.
This result is in rough agreement with a theoretical calculation of
$a_2-a_0 = (5.5\pm 0.5) a_B$ \cite{bohn:98}.
The anti--ferromagnetic interaction
energy corresponds to 2.5 nK in our condensates. Still, the
magnetostatic (ferromagnetic) interaction between the atomic magnetic
moments is about ten times weaker.
It is interesting to note that the optically trapped samples in which the
domains were observed were at a temperature of the order of 100 nK, far
larger than the anti--ferromagnetic energy. The formation of spin domains
occurs only in a \bec .

Fig. 3 {\bf c} shows a profile of the density distribution for a cloud at
$B_0=20\,{\rm mG}$ and almost canceled gradient ($B^\prime <2\,{\rm
mG/cm}$). No $m_F =0$ region can be identified. The cloud was prepared with
a small total angular momentum. Due to the almost--zero gradient and the
non--zero angular momentum the cloud corresponds to a point in the
shaded region in Fig. 1 {\bf a}, rather than a vertical line with no offset
as discussed before with finite gradients and zero angular momentum.
The different widths of the profiles are probably caused by residual
field inhomogenities. Fig. 3 {\bf c} demonstrates the complete miscibility
of the $m_F =\pm 1$ components.

For a homogenous two--component system the criterion for miscibility
(immiscibility) is 
\linebreak\mbox{$a_{ab} < (>) \sqrt{a_aa_b}$ \cite{timm:97,esry:97,ao:98}},
when the mean field energy is parametrized as $(2\pi \hbar^2/M)(n_a^2a_a
+n_b^2a_b +
2n_an_ba_{ab})$. Here, $n_{a,b}$ and $a_{a,b}$ are densities and scattering
lengths for the components $a$ and $b$, and the scattering length $a_{ab}$
characterizes the interactions between particles $a$ and $b$. In our spinor
condensate with mixtures of the $m_F\pm 1$ components, we have
$a_{-1}=a_{+1}=\bar{a}+\Delta a$ and $a_{-1\,+1}=\bar{a}-\Delta a$. Thus
$\Delta a > 0$, like experimentally observed, implies miscibility. For a
mixture of the $m_F=1$ and $m_F=0$ components, we find
$a_{0}=\bar{a}, \quad a_{+1}=\bar{a}+\Delta a$ and
$a_{0\,+1}=\bar{a}+\Delta a$, corresponding to immiscibility. For the
$^{87}$Rb experiments \cite{myat:97,hall:98a} it is not clear whether the
two components are miscible or overlap only in a surface region due to
kinetic energy \cite{corn:98}.

In conclusion,
\bec s of sodium occupying all three hyperfine states of the $F=1$ ground
state multiplet were optically trapped in low magnetic fields. The
hyperfine states are coupled by spin--exchange processes, resulting in the
formation of spin domains. 
We developed spin domain diagrams for both the anti--ferro\-magnetic and the
ferromagnetic case, and showed that sodium has anti--ferromagnetic interactions,
whereas the opposite case is predicted for the $^{87}$Rb $F=1$ spin
multiplet \cite{bohn:98}. All regions in the spin domain diagrams are accessible with our
experimental technique and thus any combination of the three hyperfine
components can be realized by applying small external magnetic fields.
Of special interest for future work is the zero magnetic field
case, where the rotational symmetry should be spontaneously broken. We observed both
miscibility and immiscibility of hyperfine components.
Thus the dynamics and possible metastable configurations \cite{law:98} of
two interpenetrating, miscible superfluid components ($m_F\!=\!\pm 1$)
with arbitrary admixtures of an immiscible component ($m_F\!=\!0$) can now
be studied.

We acknowledge stimulating discussions with Jason Ho and Chris Greene. This
work was supported by the Office of Naval Research, NSF, Joint Services
Electronics Program (ARO), NASA, and the David and Lucile Packard
 Foundation. J.S.\ would like to acknowledge support from the Alexander von
Humboldt-Foundation, D.M.S.-K. from the JSEP Graduate Fellowship Program,
and A.P.C. from the NSF.

\newpage

{\bf Figure caption 1:} Spin domain diagrams for spin--one condensates. The structure of
the ground state spinor is shown as a function of the linear ($\sim p$) and 
quadratic ($\sim q$) Zeeman energies.
Hyperfine components are mixed inside the shaded regions. Solid lines indicate
a discontinuous change of state populations whereas dashed
lines indicate a gradual change. The behaviour for $q<0$ is also shown
although it is not relevant for this experiment. For $c=0$,
the Zeeman energy causes the cloud to separate into three domains with
$m_F=+1,0,-1$ and with boundaries at $|p|=q$, as shown in {\bf b}. For $c_2\ne 0$, the
mean field energy shifts the boundary region between domains 
and leads to regions of overlapping spin components.
In the anti--ferromagnetic case ({\bf a}), the $m_F\! =\! 0$ component and
the $m_F\!=\!\pm1$ components are immiscible (including the kinetic energy terms
in Eqn. (\ref{k}) would lead to a thin boundary layer) and the
boundary occurs at $|p|=q+c$. For small bias fields, with $q<c$ and $|p|<2c$,
the $m_F\! =\! 0$ domain is bordered by domains in which $m_F\! =\! \pm 1$
components are mixed. The ratio of the $m_F\! =\! \pm 1$ populations in
these regions does not depend on $q$, but is given by $|\zeta_+|^2/|\zeta_-|^2
= (2c+p)/(2c-p)$. In this region of small fields, the boundary to the
$m_F\!=\!0$ component lies at $|p|=2\sqrt{cq}$.
In the ferromagnetic case ({\bf c}) all three components are generally
miscible, and have no sharp boundaries. Pure $m_F\!=\!0$ domains occur for
$|p|\le \sqrt{q(q-4|c|)}$ and pure $m_F\!=\!\pm 1$ domains for $|p|>q$. Here,	
in contrast to the anti--ferromagnetic case, a pure $m_F\! =\! 0$ condensate
is skirted by regions where it is mixed predominantly with either the
$m_F\!=\!-1$ or the $m_F\!=\!+1$ component. The contribution of the third
component is very small ($<2\,\%$). In all mixed regions the $m_F\!=\!0$
component is never the least populated of the three spin components. This 
qualitative feature can be used to rule out that $F=1$ sodium atoms have 
ferromagnetic interactions.

{\bf Figure caption 2:} 
Formation of ground state spin domains. Absorption images of ballistically expanding 
spinor condensates show both the spatial and hyperfine distributions.
Arbitrary populations of the three hyperfine states were prepared using rf
transitions (Landau--Zener sweeps) 
\cite{stam:98}.  At a bias field of about $40\,{\rm G}$ the transitions
from $m_F\! =\! -1$ to $m_F \! =\! 0$ and from $m_F \! =\! 0$ to
$m_F\!=\!+1$ differ in frequency by about $0.9\,{\rm MHz}$ due to the
quadratic Zeeman shift and they could be driven separately. 
The images of clouds with various dwell times
in the trap show the evolution to the same equilibrium
for condensates prepared in either a pure
$m_F =0$ state (upper row) or in equally populated $m_F =\pm1$ states
(lower row). Between $5\,{\rm s}$ to
$15\,{\rm s}$ dwell time, the distribution did not significantly change,
although the density decreased due to three--body recombination. The bias field
during the dwell time was $B_0=20\,{\rm mG}$ and the field gradient was
$B^\prime=11\,{\rm mG/cm}$.
These images were taken after the optical trap was suddenly
switched off and the atoms were allowed to expand. Due to the large aspect
ratio (typically 60), the expansion was almost purely in the radial
directions. All the mean--field energy was released after less than 1 ms, 
after which the atoms expanded as free particles.
Thus, a magnetic field gradient, which was applied after 5 ms
time--of--flight to yield a Stern--Gerlach separation of the cloud, merely
translated the three spin components without affecting their shapes. In this 
manner, the single time--of--flight images provided both a spatial and
spin--state description of the trapped cloud. Indeed, the shapes of the three
clouds fit together to form a smooth total density distribution. After a
total time--of--flight of 25 ms the atoms were optically pumped into the
$F=2$ hyperfine state and observed using the $m_F \!=\!+2$ to $m_F\!=\!+3$
cycling transition. This technique assured the same transition strength for
atoms originating from different spin states. The size of the field of view for a single
spinor condensate is 1.7 mm $\times$ 2.7 mm.

{\bf Figure caption 3:} Miscible and immiscible spin domains. Axial column density
profiles of spinor Bose-Einstein condensates are shown, obtained from time-of-flight
absorption images as in Fig. 2. The profiles of
the $m_F\!=\!\pm 1$ components were shifted to undo the Stern-Gerlach separation. 
At low bias fields (Fig. 3 {\bf a}), the $m_F\! =\! 0$ component was skirted on both
 sides by $m_F\!=\!\pm 1$ components with significant $m_F\!=\!\mp 1$
admixtures thus demonstrating the anti--ferromagnetic interaction (also
visible in Fig. 2).
At higher fields (Fig. 3 {\bf b}), the $m_F\!=\!\pm 1$ components are
pushed apart further  by a larger $m_F\!=\!0$ component and the
$m_F\!=\!\mp 1$ admixtures vanish.
They could not be resolved for quadratic Zeeman energies $q>20\,{\rm Hz}$.
The anti--ferromagnetic interaction leads to immiscibility of the
$m_F\!=\!0$ and the $m_F\!=\!\pm 1$ components. The kinetic energy in this
boundary region, which is small compared to the total mean--field energy,
is released in the axial direction. Due to this axial expansion of the
cloud in the time--of--flight
and due to imperfections in the imaging system including the limited pixel
resolution, the $m_F\! =\! 0$  to $m_F\! =\! \pm 1$ boundary is not sharp.
Fig. 3 {\bf c} demonstrates the complete miscibility
of the $m_F =\pm 1$ components. The magnetic field parameters were
$B_0=20\,{\rm mG},\,B^\prime=11\,{\rm mG/cm}$ in {\bf a},
$B_0=100\,{\rm mG},\,B^\prime=11\,{\rm mG/cm}$ in {\bf b}, and
$B_0=20\,{\rm mG},\,B^\prime <2\,{\rm mG/cm}$ in {\bf c}.

{\bf Figure caption 4:} 
Estimate of the anti--ferromagnetic interaction energy $c$. Plotted is the linear
Zeeman energy $p = |\mu B^\prime z_b|$ at the boundary between the
$m_F\!=\!0$ and $m_F\!=\!\pm1$ regions versus the quadratic Zeeman
shift $q=\hat{q}B_0^2$.
$\hat{q} = g_s^2\mu_B^2/16 h^2\nu_{hfs} = 278\, {\rm Hz/G^2}$, and $\mu
 = g_s\mu_B /4h=  700 \,{\rm kHz/G}$, $g_s$ denotes the electron
g--factor, $\nu_{hfs}$ the hyperfine splitting frequency, and $\mu_B$ the
Bohr magneton. The solid line is a fit of the function $|p|=2\sqrt{qc}$ for $q<c$ and
$|p|=q+c$ for $q>c$. Extrapolating the linear part to zero bias field (dashed line) yields
$c=(50\pm 20)\,{\rm Hz}$.
The data points at a given bias field represent $p = \mu B^\prime z_b$ for
different gradient fields $B^\prime$ and thus $m_F\!=\!0$ regions of
different size. The scatter of these points is mainly due to a residual
magnetic field inhomogenities, resulting in small deviations of the local gradient
$B^\prime$. The error bar represents the relative error of all data points of 
30 \% in $p$ and 5 \% in $q$ as estimated from the uncertainties in the magnetic
field calibration. Furthermore the limited pixel resolution and contributions of the
kinetic energy in the condensate to the axial expansion
enhance the errors for the determination of $z_b$ of small $m_F =0$ regions.


\clearpage
\newpage

\begin{figure}[htbf]
\unitlength1cm
\begin{picture}(15,6)\epsfxsize=100mm
 \centerline{\epsffile{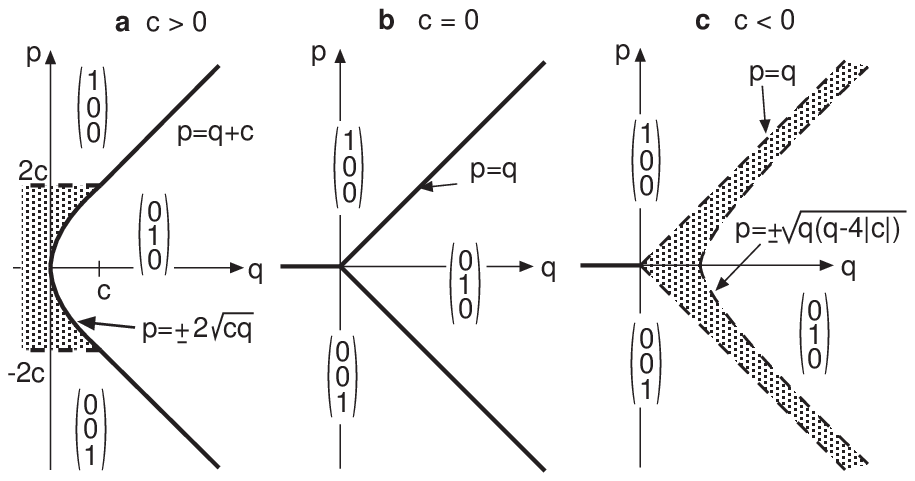}}
\end{picture}
\caption{}
\label{fig1}
\end{figure}


\begin{figure}[htbf]
\unitlength1cm
\begin{picture}(15,10)\epsfxsize=80mm
 \centerline{\epsffile{fig2.epsf}}
\end{picture}
\caption{}
\label{fig2}
\end{figure}


\begin{figure}[htbf]
\unitlength1cm
\begin{picture}(10,9)\epsfxsize=75mm
 \centerline{\epsffile{fig3.epsf}}
\end{picture}
\caption{}
\label{fig3}
\end{figure}

\begin{figure}[htbf]
\unitlength1cm
\begin{picture}(10,7)\epsfxsize=80mm
 \centerline{\epsffile{fig4.epsf}}
\end{picture}
\caption{}
\label{fig4}
\end{figure}

\end{document}